\documentclass[slac_one]{revtex4}
\usepackage{graphicx}
\usepackage{fancyhdr}
\usepackage{xspace}
\newcommand{\W}{\ensuremath{W}\xspace}

\newcommand{\Z}{\ensuremath{Z}\xspace}
\newcommand{\gev}{\ensuremath{\mathrm{\,Ge\kern -0.1em V}}\xspace}
\newcommand{\mev}{\ensuremath{\mathrm{\,Me\kern -0.1em V}}\xspace}

\newcommand{\ptl}{\ensuremath{p_T^{\ell}}\xspace}

\newcommand{\ptn}{\ensuremath{p_T^\nu}\xspace}
\newcommand{\ptw}{\ensuremath{p_T^W}\xspace}

\newcommand{\mtw}{\ensuremath{m_T^W}\xspace}

\newcommand{\phil}{\ensuremath{\phi_{\ell}}\xspace}
\newcommand{\phin}{\ensuremath{\phi_{\nu}}\xspace}

\newcommand{\yw}{\ensuremath{y_W}\xspace}
\newcommand{\yz}{\ensuremath{y_Z}\xspace}

\newcommand{\met}{\ensuremath{E_T \hspace*{-3.0ex} \not~~}\xspace}

\newcommand{\invpb}{\ensuremath{\mbox{\,pb}^{-1}}\xspace}

\newcommand{\invfb}{\ensuremath{\mbox{\,fb}^{-1}}\xspace}

\begin{document}

\title{Measurement of the W mass with the ATLAS detector}

\author{T. C. Petersen (for the ATLAS collaboration)}
\affiliation{CERN, Geneve, Switzerland}

\begin{abstract}
  We investigate the posibility of improving the \W mass measurement
  at ATLAS. Given the high statistics of both \W and \Z bosons
  expected at the LHC, we estimate that a precision of 7 \mev per
  channel can be reached with $10\,\invfb$.
\end{abstract}

\maketitle

\thispagestyle{fancy}

\section{Motivation}

\noindent
The Standard Model (SM) is a very predictive framework. Given precise
measurements of $\alpha_{QED}$, $G_\mu$, and $m_Z$, the $W$ boson mass
plays a central role, as it allows for both a SM cross check, confronting
predictions of the $W$ and top quark masses~\cite{:2005ema} with
measurements~\cite{Abe:1995hr,Abachi:1995iq}, and limits on the SM
Higgs boson mass~\cite{Alcaraz:2006mx}. Finally, constraints on the
contributions of other heavy particles, like supersymmetric
particles~\cite{Heinemeyer:2004gx} can be obtained.
The $W$ mass precision has continually improved with statistics,
yielding the current world average of $m_W = 80.398 \pm 0.025
\gev$~\cite{:2007ypa}. Further improvement will translate into
more precise indirect predictions of the SM Higgs mass.

\section{Event selection}

\noindent
The simulated $W$ and $Z$ boson signal and associated background
samples used in this study are computed using the {\tt PYTHIA} general
purpose event generator~\cite{Sjostrand:2006za}, with photon radiation
in \W and \Z decays treated via an interface to {\tt
PHOTOS}~\cite{Golonka:2005pn}.  The size of the expected samples are
computed assuming the NLO \W and \Z cross-sections, as obtained from
{\tt RESBOS}~\cite{Balazs:1997xd}, and simulated with complete
simulation of the ATLAS detector using {\tt GEANT4}~\cite{Rimoldi:2005tg}.

\noindent
At hadron colliders, \W and \Z events can be detected and
reconstructed in the $e\nu_e$, $\mu\nu_\mu$, $ee$, and $\mu\mu$ final
states. In the following, the term lepton ($\ell$) will refer to
either an electron or muon.
Electrons are measured using the inner detector (ID) and
electromagnetic calorimeter (EMC). They are reconstructed and
identified with an efficiency of about 65\%, while rejecting
background from jets up to one part in $10^5$. The transition
region from barrel to endcap in the EMC ($1.3 < |\eta| < 1.6$) is not
used.
For muons, the ID is used together with the muon spectrometer with a
reconstruction efficiency of about 95\%. Backgrounds are less than for
electrons, and diminished using isolation.
The transverse momentum of the neutrino is inferred from the
transverse energy imbalance as determined by the calorimeters.
The relative energy resolution is typically 1.5\% for electrons and
2.0\% for muons, while the missing transverse momentum (MET) has a
resolution of 15-25\%~\cite{:2008zzm}.

\noindent
The \W signal is extracted by selecting events with one isolated
lepton ($\ptl > 20\,\gev$ and $|\eta_\ell| < 2.5$) along with
significant MET due to the undetected neutrino ($\met > 20\,\gev$).
These selections have a total efficiency (trigger and selection) of
about 20\% (40\%) for the electron (muon) channel, providing a sample
of about $4 \times 10^7$ ($8 \times 10^7$) events. The backgrounds
are at the 3\% (6\%) percent level.
Likewise, the \Z signal is required to have two opposite sign leptons
($\ptl > 20\,\gev$ and $|\eta_\ell|<2.5$). The efficiency of this
selection is about 10\% (30\%) in the electron (muon) channel,
yielding samples of about $2 \times 10^6$ ($7 \times 10^6$) events.
\begin{table}[bhp!]
\begin{center}
\begin{tabular}{lcccc}
  \hline
  \hline
    Channel                  &$W \to e\nu$        &$W \to \mu\nu$          &$Z \to ee$           &$Z \to \mu\mu$\\
  \hline
    Reconstructed lepton(s)  &\multicolumn{4}{c}{$p_T > 20 \gev$, $|\eta| < 2.5$}\\
    Crack region removed     &$1.30 < |\eta| < 1.60$    &--                &$1.30 < |\eta| < 1.60$      &--\\
    Missing energy           &\multicolumn{2}{c}{$\met > 20 \gev$}         &--                   &--\\
  \hline
    Events in 10 \invfb\ [$10^6$]   & 47         & 84           &  2.1     &  6.7\\
  \hline
  \hline
\end{tabular}
\caption{Selection criteria for \W and \Z events in electron and muon
  channel, and resulting statistics for 10~\invfb\ of data.
  \label{tab:selection_stat}}
\end{center}
\end{table}

\noindent
While the invariant mass can be determined in \Z boson events, the
observables most sensitive to $m_W$ are:
\begin{itemize}
  \item The reconstructed lepton transverse momentum, \ptl.
  \item The reconstructed \W transverse mass,
  $\mtw \equiv \sqrt{2 \ptl \ptn (1 - \cos(\phil - \phin))}$.
\end{itemize}
Based on the \ptl\ and \mtw\ distributions, $m_W$ can be extracted by
comparing the data to a set of models (template distributions)
obtained by varying the value of the \W boson mass parameter in the
event generation. With 10~fb$^{-1}$ of data the statistical precision
is about $2\,\mev$ for each channel, roughly matching that of the
smaller but more precise \Z samples.

\noindent
For the above procedure to work in practice, one must predict the
\ptl\ and \mtw\ distributions as a function of the \W mass. These
distributions are however affected by many effects, which need to be
included correctly in order to avoid biases in the mass fit.
%
%
%
The impact of mechanisms affecting the \W mass determination is
estimated by producing template distributions of \ptl\ and \mtw\
\emph{unaware of the effect} under consideration, and fitting them to
distributions \emph{including this effect}. The resulting bias yields
the corresponding systematic uncertainty.

\section{Calibration and experimental uncertainties}
\label{sec:exp_unc}

\begin{figure*}[htbp!]
  \centering
  \includegraphics[width=91mm]{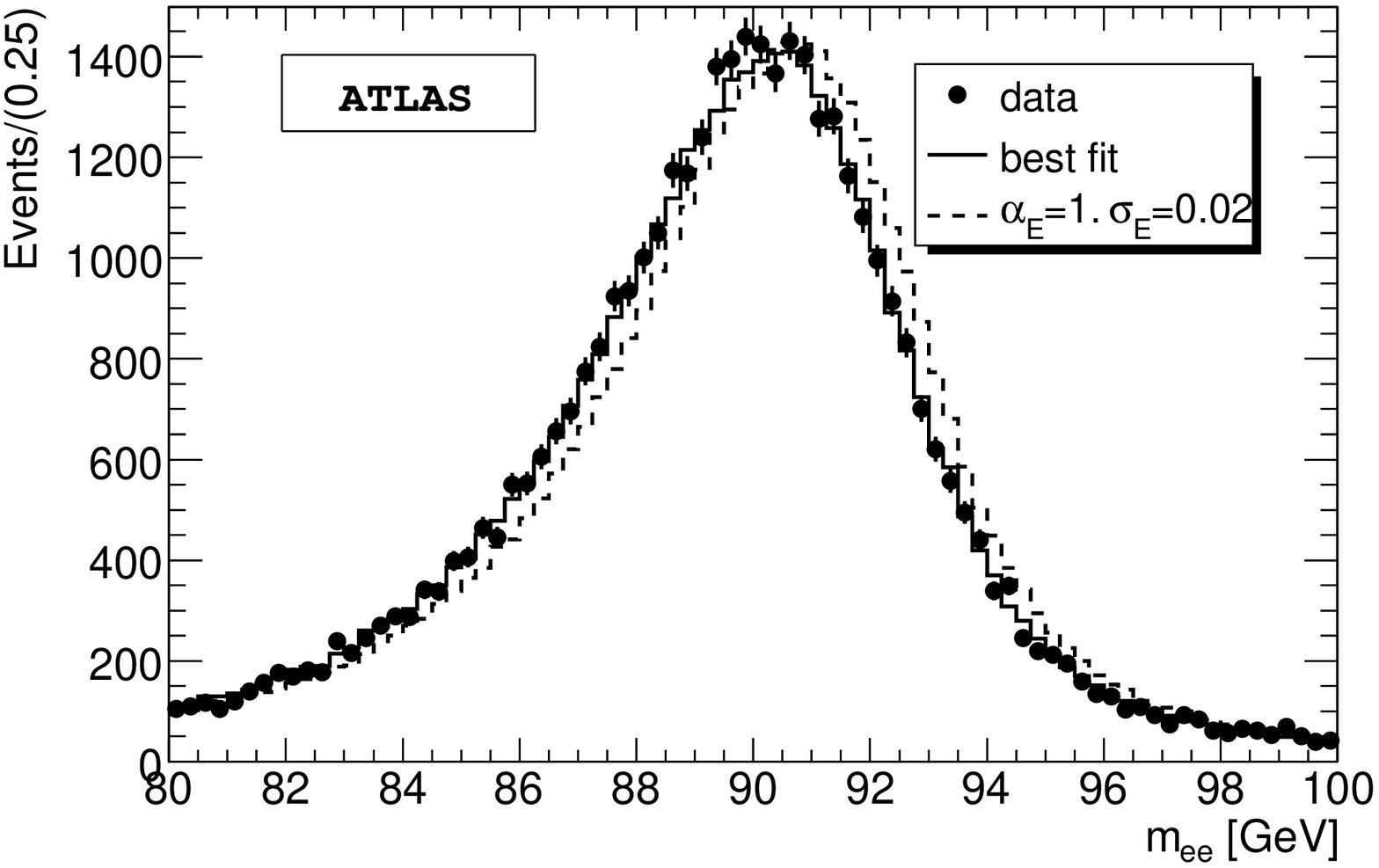}
  \includegraphics[width=86mm]{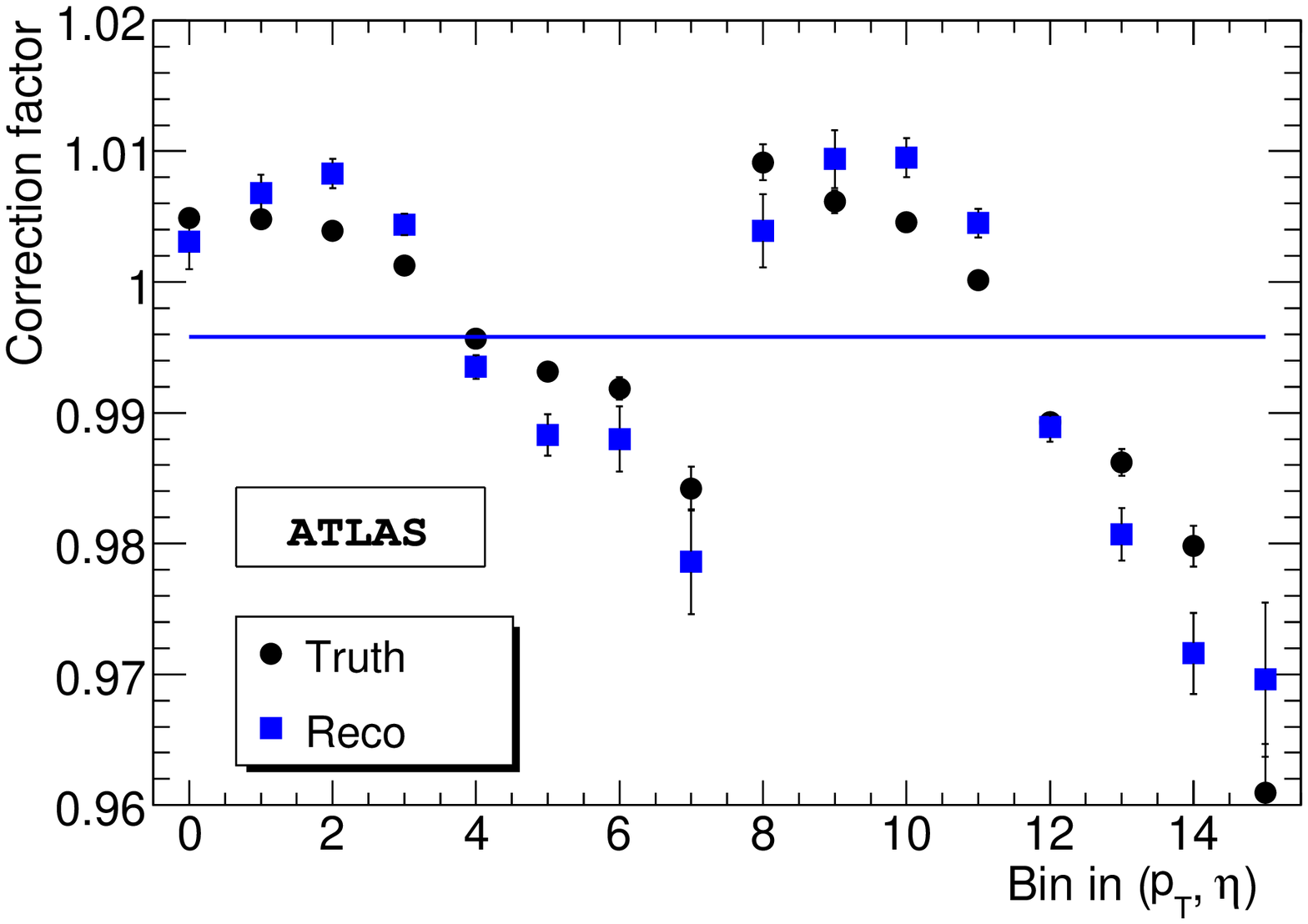}
\caption{{\bf Left:} Average calibration using $3.5 \times 10^5$ $Z
  \to ee$ events. {\bf Right:} Differential (linearity) calibration as
  a function of $p_T$ (8 bins) and $\eta$ (2 bins) with same data. The
  two calibrations agree with each other and the generated value.
  \label{fig:calibration_scale}}
\end{figure*}

\noindent
The precise knowledge of the \Z mass and width \cite{:2005ema} allows
for an accurate determination of the lepton energy scale and
resolution.
Given a sample of 30700 reconstructed $Z \to ee$ events (${\cal L} \sim
100\,\invpb$) with $85 < m_{ee} < 97\,\gev$, an average mass
scale (defined as $\alpha = m_Z^{\mbox{\tiny reco}}
m_Z^{\mbox{\tiny truth}}$) of $\alpha = 0.9958 \pm 0.0003$ was
obtained on a fully simulated example sample (see
Figure~\ref{fig:calibration_scale} left).\\
In order to correctly propagate the \Z calibration measurement to the
\W sample, the scale needs to be measured as a function of energy.
The high statistics expected at LHC allows for the refinement of doing
the above calibration differentially in $p_T$ and $\eta$ (here in $8
\times 2$ bins), exploiting the energy distribution of the decay
leptons, and hence measuring the linearity of the detector response.\\
Each event is assigned to a category $(i,j)$, according to $p_T \times
\eta$ bins (16 in total) of the two leptons (choosing $i \geq j$).
For each category $(i,j)$, the reconstructed sample is compared to the
known \Z lineshape, and a \Z mass resolution function $R_{ij}$ is
obtained from requiring that its convolution with the theoretical
lineshape matches the reconstructed distribution.
The \Z mass resolutions $R_{ij}$ result from combining two lepton
momentum resolutions $R_i$ and $R_j$ as $R_{ij} = R_i \otimes R_j$.
Given $N$ lepton bins and thus lepton resolution functions to determine,
there are $N \times (N+1) / 2$ \Z mass resolution functions, and thus
the overconstrained system can be solved by a global $\chi^2$ fit,
allowing for a determination of the detector response for all
combinations of $p_T$ and $\eta$ (see
Figure~\ref{fig:calibration_scale} right).

\noindent
Once the lepton scale is established, the \Z transverse momentum will
also serve to scale the measured hadronic recoil to the \Z, which
together with the measured lepton transverse momentum defines the
missing transverse energy. Finally, ``tag and probe''
methods~\cite{Abbott:1999wk} will allow to determine the lepton
reconstruction efficiency.

\noindent
Backgrounds are small and mostly from well known similar heavy boson decays
yielding true leptons (estimated from simulation), or from dijet
events (estimated using two independent discriminators) faking leptons.

\section{Theoretical uncertainties}
\label{sec:theo_unc}

\noindent
Most QCD mechanisms affecting \W distributions carry significant
uncertainty, but affect \W and \Z events in a similar way.
This is the case for non-perturbative contributions, but also for
parton density (PDF) effects. At the LHC, the \W and the \Z are
essentially sensitive to high-$Q^2$ sea partons, and a variation of
these parameters will affect the \W and \Z distributions (in
particular \yw and \yz) in a highly correlated way.
Since the usage of the \Z for calibration effectively makes the
analysis a measurement of the \W to \Z mass ratio, the impact of
correlated effects is strongly constrained.
Evaluation was based on variation of parameters.

\noindent
The \W width uncertainty was assumed to diminish at the LHC.
The impact of QED radiation was evaluated by varying the order of the
QED calculation by PHOTOS and considering general LEP precision.
For details, see \cite{Besson:2008zs}.

\section{Results and Conclusion}

\noindent
The systematic uncertainties are summarized in Table~\ref{tab:results}
for $10\,\invfb$ of data. Assuming expected detector performance and
required theoretical tools to be available, the result is a precision
on $m_W$ of 7 \mev per channel.
Additional calibration processes and combining independent
measurements may bring further improvement.
\begin{table}[htp!]
\begin{center}
\begin{tabular}{l@{\hspace*{3ex}}c@{\hspace*{3ex}}c@{\hspace*{7ex}}l@{\hspace*{3ex}}c@{\hspace*{3ex}}c}
\hline
\hline
  Experimental effect           &$\sigma(m_W)$ (\ptl)
                                      &$\sigma(m_W)$ (\mtw)
                 &Theoretical effect           &$\sigma(m_W)$ (\ptl)
                                                     &$\sigma(m_W)$ (\mtw)\\
\hline
  Lepton scale, lin.\ \& res.\  &4    &4                       
                 &\W width                     &0.5  &1.3\\
  Lepton efficiency             &4.5 ($e$), $<1$ ($\mu$)   &4.5 ($e$), $<1$ ($\mu$)
                 &\yw\ distribution            &1    &1\\
  Recoil scale, lin.\ \& res.\  &--   &5
                 &\ptw\ distribution           &3    &1\\
  Bkg.\ (heavy bosons)          &2    &1.5
                 &QED radiation                &$<$1 &$<$1\\
  Bkg.\ (dijets)                &0.5  &0.4
                 &                             &     &\\
\hline
                               &          &        
                 &{\bf Total}                  &$\sim$7 (e); 6 ($\mu$)  &$\sim$8 (e); 7 ($\mu$) \\
\hline
\hline
\end{tabular}
\caption{Breakdown of systematic uncertainties for \ptl and \mtw fits
  in $e$ and $\mu$ channel to $m_W$. All number are in \mev.
  \label{tab:results}}
\end{center}
\vspace*{-3ex}
\end{table}


%
%


\small
\begin{thebibliography}{14}   

\bibitem{:2005ema}
    [ALEPH Collab., DELPHI Collab., L3 Collab., ],
  Phys.\ Rept.\  {\bf 427} (2006) 257
  [arXiv:hep-ex/0509008].

\bibitem{Abachi:1995iq}
  S.~Abachi {\it et al.}  [D0 Collaboration],
  Phys.\ Rev.\ Lett.\  {\bf 74} (1995) 2632
  [arXiv:hep-ex/9503003].

\bibitem{Abe:1995hr}
  F.~Abe {\it et al.}  [CDF Collaboration],
  Phys.\ Rev.\ Lett.\  {\bf 74} (1995) 2626
  [arXiv:hep-ex/9503002].

\bibitem{Alcaraz:2006mx}
  J.~Alcaraz {\it et al.}  [ALEPH Collab., DELPHI Collab., L3 Collab., ],
  arXiv:hep-ex/0612034.

\bibitem{Heinemeyer:2004gx}
  S.~Heinemeyer, W.~Hollik and G.~Weiglein,
  Phys.\ Rept.\  {\bf 425} (2006) 265
  [arXiv:hep-ph/0412214].

\bibitem{:2007ypa}
  T.~Aaltonen {\it et al.}  [CDF Collaboration],
  Phys.\ Rev.\ Lett.\  {\bf 99} (2007) 151801
  [arXiv:0707.0085 [hep-ex]].

\bibitem{Sjostrand:2006za}
  T.~Sjostrand, S.~Mrenna and P.~Skands,
  JHEP {\bf 0605} (2006) 026
  [arXiv:hep-ph/0603175].

\bibitem{Golonka:2005pn}
  P.~Golonka and Z.~Was,
  Eur.\ Phys.\ J.\  C {\bf 45} (2006) 97
  [arXiv:hep-ph/0506026].

\bibitem{Balazs:1997xd}
  C.~Balazs and C.~P.~Yuan,
  Phys.\ Rev.\  D {\bf 56} (1997) 5558
  [arXiv:hep-ph/9704258].

\bibitem{Rimoldi:2005tg}
  A.~Rimoldi {\it et al.},
{\it Prepared for 9th ICATPP Conference, Villa Erba, Como, Italy, 17-21 Oct 2005}

\bibitem{:2008zzm}
  G.~Aad {\it et al.}  [ATLAS Collaboration],
  JINST {\bf 3} (2008) S08003

\bibitem{Abbott:1999wk}
  B.~Abbott {\it et al.}  [D0 Collaboration],
  Phys.\ Rev.\  D {\bf 61} (2000) 032004
  [arXiv:hep-ex/9907009].

\bibitem{Berge:2005nm}
  S.~Berge, P.~M.~Nadolsky, F.~I.~Olness and C.~P.~Yuan,
  AIP Conf.\ Proc.\  {\bf 792} (2005) 722
  [arXiv:hep-ph/0508215].

\bibitem{Besson:2008zs}
  N.~Besson, M.~Boonekamp, E.~Klinkby, T.~Petersen and S.~Mehlhase,
  arXiv:0805.2093 [hep-ex].

\end{thebibliography}
\end{document}